\documentclass[letterpaper, 10 pt, conference]{ieeeconf}  

\IEEEoverridecommandlockouts                              
\usepackage{amsmath,amssymb,amsfonts}
\usepackage{graphicx}
\usepackage{tikz}
\usepackage{pgfplots}
\pgfplotsset{compat=1.9}
\usepackage{pgfplotstable}
\usepgfplotslibrary{groupplots}
\usepackage[disable]{todonotes} 
\usepackage{array} 
\usepackage{booktabs}
\usepackage[font=small,skip=0.3pt]{caption}
\usepackage{hyperref}

\usepackage{algorithm,algorithmic}
\let\labelindent\relax
\usepackage{enumitem} 

\newcommand{\jc}[1]{\todo[inline,color=white!50!green]{#1}}

\DeclareMathOperator*{\argmin}{arg\,min}

\newtheorem{assumption}{\textbf{Assumption}}

\newtheorem{lemma}{\textbf{Lemma}}
\newtheorem{corollary}[lemma]{\textbf{Corollary}}

\newtheorem{remark}{\textbf{Remark}}

\newcommand{\change}[1]{\textcolor{black}{#1}}

\newcommand{\changeE}[1]{\textcolor{black}{#1}}

\title{\LARGE \bf
Efficient Recursive Data-enabled Predictive Control (Extended Version)
}

\author{Jicheng Shi, Yingzhao Lian and Colin N. Jones
\thanks{This work received support from the Swiss National Science Foundation (SNSF) under the NCCR Automation project, grant agreement 51NF40\_180545. (corresponding author: Jicheng Shi)
}
\thanks{JS, YL and CNJ are with Automatic Laboratory, EPFL, 1015 Lausanne, Switzerland. {\tt\small $\{$jicheng.shi, yingzhao.lian, colin.jones$\}$@epfl.ch}}\vspace{-1em}}

\begin{document}

\maketitle


\begin{abstract}
In the field of model predictive control, Data-enabled Predictive Control (DeePC) offers direct predictive control, bypassing traditional modeling. However, challenges emerge with increased computational demand due to recursive data updates. This paper introduces a novel recursive updating algorithm for DeePC. It emphasizes the use of Singular Value Decomposition (SVD) for efficient low-dimensional transformations of DeePC in its general form, as well as a fast SVD update scheme. 
Importantly, our proposed algorithm is highly flexible due to its reliance on the general form of DeePC, which is demonstrated to encompass various data-driven methods that utilize Pseudoinverse and Hankel matrices. This is exemplified through a comparison to Subspace Predictive Control, where the algorithm achieves asymptotically consistent prediction for stochastic linear time-invariant systems. 
Our proposed methodologies' efficacy is validated through simulation studies.
\end{abstract}

\jc{consistent prediction for SLTI systems\\
Data-Driven Predictive Control With Online Adaption: Application to a Fuel Cell System}

\section{Introduction}\label{sect:intro}
In Model Predictive Control (MPC), data-driven techniques have emerged as promising tools to expedite and enhance controller design, offering end-to-end solutions from input-output (I/O) data to fully functional controllers. Among these, Data-enabled Predictive Controller (DeePC) has gained significant attention, leveraging Willems' Fundamental Lemma~\cite{willems2005note} to bypass traditional modeling steps and establish a direct predictive controller. This method has demonstrated effectiveness across diverse domains, including batteries~\cite{schmitt2023data,chen2022data}, buildings~\cite{shi2023adaptive,yin2024data}, grids~\cite{huang2019data}, and vehicles~\cite{wang2023implementation}.

In deterministic Linear Time-invariant (LTI) systems, numerous data-driven approaches have demonstrated the capability to consistently estimate system dynamics using a finite yet sufficiently excited I/O dataset. A paradigmatic method in this regard, Subspace Identification (SID)~\cite{SIDvan2012subspace}, employs an indirect approach to generate a consistent state-space model, facilitating the subsequent design of an MPC \change{controller}. In addition, other direct methods such as DeePC and Subspace Predictive Control (SPC)~\cite{SPCfavoreel1999spc} can leverage this limited data set to directly yield \change{exact} trajectory prediction.

The landscape changes slightly for LTI systems affected by stochastic noise. Several data-driven studies have shown that employing infinite open-loop I/O data leads to the estimation of asymptotically consistent models~\cite{qin2006overview}. Building on this foundation, more recent efforts have sought to extend these algorithms to closed-loop data~\cite{van2013closed}. An innovative proposal in this context, as mentioned in~\cite{CLSPCdong2008closed}, seeks to design the SPC using initial I/O data for system control, with the SPC undergoing recursive updates to enhance performance. DeePC has also witnessed similar extensions, such as the integration of instrumental variables~\cite{OPDEEPCvan2022data, CLDEEPCdinkla2023closed, wang2023data}, which have also been explored to achieve consistency in predictions utilizing both open~\cite{OPDEEPCvan2022data} and closed-loop data~\cite{CLDEEPCdinkla2023closed}.

A major hurdle arises with DeePC's increasing computational complexity as more I/O data are integrated. Recent studies have aimed to mitigate this by reducing DeePC's computational overhead~\cite{chen2022data,LOWDEEPCzhang2022dimension,baros2022online}. For instance, \cite{chen2022data,LOWDEEPCzhang2022dimension} employ the Singular Value Decomposition (SVD) of the Hankel matrix to reduce the dimensions of DeePC's decision variables. However, these methods, while promising, often present their own challenges, especially as more data is recursively incorporated into the model. Existing recursive updating methodologies in SID~\cite{lovera2000recursive} and SPC~\cite{CLSPCdong2008closed,verheijen2022recursive}, reliant on the least squares structure, remain unsuited to DeePC and its variations. This gap underscores the pressing need for a generalized and efficient strategy to recursively update DeePC.

Addressing this void, our research introduces an effective recursive updating paradigm within the DeePC framework. Our main contribution \change{describes} the equivalency between \change{an} SVD-based low-dimensional DeePC and its counterpart in a more general form compared to~\cite{chen2022data,LOWDEEPCzhang2022dimension}, while also detailing an efficient SVD updating mechanism for recursively updated I/O data. 
A key advantage of the proposed algorithm is its high degree of flexibility rooted in the general-form DeePC. 
Our study demonstrates that this form of DeePC can include data-driven methods based on Pseudoinverse and Hankel matrices.
We give an example of this through a comparison to SPC, where the algorithm achieves asymptotically consistent prediction.
In addition, our proposed algorithm has the potential for broader applications, especially among other adaptive DeePC methods ~\cite{lian2021adaptive,berberich2022linear}.

The paper's structure is as follows: Section~\ref{sect:2} revisits Willems' fundamental lemma and DeePC. Subsequently, Section~\ref{sect:3} delves into the equivalent low-dimensional transformation of DeePC, introducing our efficient recursive updating method. Section~\ref{sect:4} details an extension to data-driven methods based on Pseudoinverse. The validity of these methods is empirically established through simulations presented in Section~\ref{sect:5}.


\noindent \textbf{Notation:} 
Let $\mathbf{0}$ represent a zero matrix, and $I$ represent an identity matrix. The notation $x: = \{x_i\}_{i=1}^T$ indicates a sequence of size $T$. The term $x_t$ represents the measurement of $x$ at the instance $t$. Additionally, $x_{1:L}:=[x_1^\top,x_2^\top,\ldots,x_L^\top]^\top$ signifies a concatenated sequence of $x$ from $x_1$ to $x_L$. $M^{\dag}$ indicates Moore–Penrose inverse of a matrix $M$.



\vspace{-.3em}
\section{Preliminaries}\label{sect:2}

Consider a  linear time-invariant (LTI) system described by the equations $x_{t+1} = Ax_t + Bu_t$ and $y_t = Cx_t + Du_t$,  which we refer to as $\mathfrak{B}(A,B,C,D)$. The system's order is given by $n_x$ \change{with $n_u,\, n_y$ denoting its input and output dimensions}. An $L$-step trajectory for this system is represented as $\begin{bmatrix}u_{1:L}^\top & y_{1:L}^\top\end{bmatrix}^\top$.
The set of all potential $L$-step trajectories produced by $\mathfrak{B}(A,B,C,D)$ is denoted by $\mathfrak{B}_L(A,B,C,D)$. We define the Hankel matrix $H_s$ of depth $L$ associated with a vector-valued signal sequence $s=\{s_i\}_{i=1}^T$ as:
\begin{align*}
    H_s :=
    \begin{bmatrix}
    s_1 & s_2&\dots&s_{T-L+1}\\
    s_2 & s_3&\dots&s_{T-L+2}\\
    \vdots &\vdots&&\vdots\\
    s_{L} & s_{L+1}&\dots&s_T
    \end{bmatrix}\;.
\end{align*}
The row and column counts of a Hankel matrix $H$ are given by $\textit{row}_{H}$ and $\textit{col}_{H}$, respectively. Throughout this paper, the term $L$ is exclusively used to indicate the size of the Hankel matrix. An input measurement sequence defined as $u=\{u_i\}_{i=1}^T$ is termed \textit{persistently exciting} of order $L$ if the Hankel matrix $H_u$ has full row rank.





Utilizing the Hankel matrices $H_u$ and $H_y$, we introduce the well-established \textbf{Willems' Fundamental Lemma}:

\begin{lemma}\label{lemma:funda}\cite[Theorem 1]{willems2005note}
Given a controllable linear system where $\{u_i\}_{i=1}^T$ is persistently exciting of order $L+ n_x$, the condition $\text{colspan}(\begin{bmatrix} H_u^\top & H_y^\top \end{bmatrix}^\top) = \mathfrak{B}_L(A,B,C,D)$ is satisfied.
\end{lemma}


Recent advancements in the data-driven control domain have given rise to schemes like DeePC~\cite{coulson2019data}, along with numerous variants, for instance, \cite{coulson2019regularized,lian2021adaptive}. Lemma~\ref{lemma:funda} plays a pivotal role in these schemes by facilitating trajectory prediction.
In the scope of this paper, our primary aim is to unveil a universally efficient updating algorithm tailored for various controllers under the DeePC paradigm. To exemplify, consider the L2 regularized DeePC (L2-DeePC) detailed in~\cite{coulson2019regularized}:
\begin{subequations}\label{eqn:deepc_L2}
\begin{align}
         \min_{\substack{g,\sigma\\u_{pred},y_{pred}}} \; & J(u_{pred},y_{pred}) + \lambda_{\sigma}\lVert\sigma\rVert_2^2 + \lambda_g \lVert g\rVert_2^2 \label{eqn:deeepc_L2_obj}\\
        \text{s.t.}\; &Hg=\begin{bmatrix}
        y_{init}+\sigma\\y_{pred}\\u_{init}\\u_{pred}
    \end{bmatrix}, \\
    & y_{pred} \in \mathbb{Y}, u_{pred} \in \mathbb{U} \label{eqn:deeepc_L2_cons}
\end{align}
\end{subequations}
where $H:=\begin{bmatrix}
        H_y \\ H_u
\end{bmatrix}$ for simplification.  The parameters $\lambda_{\sigma}$ and $\lambda_g$ represent user-determined regularization cost weights. The elements $J(u_{pred},y_{pred})$, $\mathbb{Y}$, and $\mathbb{U}$ are defined according to the task at hand. Sequences $u_{init}$ and $y_{init}$ provide $n_{init}$-step historical data for measured inputs and outputs leading up to the present moment, which aids in current state estimation of the dynamic system~\cite{coulson2019data}. Correspondingly, $u_{pred}$ and $y_{pred}$ denote the predicted sequences of $n_{pred}$ steps from the current timestamp. Consistently, the row dimension of the Hankel matrix is set to $L=n_{init}+n_{pred}$.



The L2-DeePC as presented in \eqref{eqn:deepc_L2} forecasts the $n_{pred}$-step output trajectory $y_{pred}$ based on a provided predictive input sequence $u_{pred}$. The objective, specified in~\eqref{eqn:deeepc_L2_obj}, is minimized subject to the constraint delineated in \eqref{eqn:deeepc_L2_cons}. The inclusion of the slack variable $\sigma$ ensures feasibility for L2-DeePC. Meanwhile, regularization terms are introduced to enhance predictions, especially beneficial when the system is prone to noise or embodies nonlinear elements. For an in-depth discussion and detailed insights, readers are directed to~\cite{coulson2019regularized}.


This paper introduces a data-driven MPC technique under the DeePC framework that is recursively updated with the most recent operational data. We term this approach recursive DeePC and detail it in Algorithm~\ref{alg:Rdeepc}. 

\begin{algorithm} 
  \caption{Recursive DeePC}
  \label{alg:Rdeepc} 
{
\begin{itemize} [nolistsep,leftmargin=1.5em]
    \item[0)] Retrieve some persistently excited past I/O data and build the initial DeePC controller, such as~\eqref{eqn:deepc_L2}.
    \item[1)] Retrieve the recent $L$-step measurements and update the Hankel matrix as:
    \begin{align} \label{eqn:recurive_update}
        H \leftarrow 
        \begin{bmatrix}
        H &
        \begin{bmatrix}
        y_{t-L:t-1}\\ u_{t-L:t-1}
        \end{bmatrix}      
        \end{bmatrix}  
    \end{align}
    \item[2)] Retrieve the recent $t_{init}$-step measurements. Solve the DeePC and apply the optimal input as $u_t=u_{pred}^{\ast}(1)$.
    \item[3)] Pause until the subsequent sampling time, update $t\leftarrow t+1$ and revert to step 1.
\end{itemize}
}
\end{algorithm}
{\footnotesize \raggedleft $u_{pred}^{\ast}(1)$ represents the first optimal input in $u_{pred}^{\ast}$ \par}

\change{
In many applications, empirical evidence suggests that DeePC benefits from larger quantities of I/O data beyond the minimal requirement~\cite{shi2023adaptive,huang2019data}. Furthermore, research in~\cite{OPDEEPCvan2022data} establishes that infinite open-loop data can ensure consistent prediction in DeePC methods with instrumental variables for stochastic LTI systems. This insight can be broadened to closed-loop data, leveraging techniques from SID~\cite{SIDvan2012subspace} and SPC~\cite{CLSPCdong2008closed}. Notably, by modifying~\eqref{eqn:recurive_update} to incorporate a forgetting factor or discard outdated data, Algorithm~\ref{alg:Rdeepc} can be adapted for adaptive DeePC approaches such as those detailed in~\cite{lian2021adaptive,berberich2022linear}.
}

\section{Efficient recursive updates in the DeePC framework }\label{sect:3}

In this section, we introduce a more computationally efficient version of Algorithm~\ref{alg:Rdeepc} when the DeePC used is in some general form. This improved algorithm relies on two primary components: (1) an equivalent low-dimensional transformation of the DeePC in the general form leveraging SVD, and (2) a fast SVD updating technique. The complete methodology is concluded in Algorithm~\ref{alg:ERdeepc}. To detail its operation, we reference the L2-DeePC~\eqref{eqn:deepc_L2} as a demonstrative example.

\subsection{An equivalent low-dimensional transformation}

For the first component of Algorithm~\ref{alg:ERdeepc}, we \change{describe} an equivalent low-dimensional transformation of a general DeePC problem. This transformation is facilitated by the SVD of the aggregated Hankel matrix, $H$:
\begin{align*}
    & H = \begin{bmatrix}U_1 & U_2\end{bmatrix} \begin{bmatrix}
        \Sigma & \mathbf{0} \\
        \mathbf{0} & \mathbf{0}
    \end{bmatrix}
    \begin{bmatrix}V_1 & V_2\end{bmatrix}^{\top} 
    = U_1 \Sigma V_1^{\top}
\end{align*}
where $\Sigma \in \mathbb{R}_{r_{H}, r_{H}}$ and $r_{H}$ is the rank of $H$. A general DeePC problem is defined as: \\
\textbf{Problem 1:} 
\begin{equation}\label{eqn:deepc_general}
\begin{aligned}
    \min_{\substack{g,\sigma\\u_{pred},y_{pred}}} \; & f_1(u_{pred},y_{pred},\sigma,V_1^{\top} g) +  f_2(V_2^{\top} g) \\
    \text{s.t.}\; &Hg=\begin{bmatrix}
        y_{init}\\y_{pred}\\u_{init}\\u_{pred}
    \end{bmatrix}+\sigma, \\
    & f_3(y_{pred}, u_{pred}, \sigma) \leq 0
\end{aligned}
\end{equation}
Here, functions $f_1(\cdot)$, $f_2(\cdot)$, and $f_3(\cdot)$ are user-specified and vary across different DeePC methodologies tailored for diverse applications, which can cover the L2-DeePC and various DeePC variants~\cite{coulson2019regularized,huang2019data,lian2021adaptive}. The aforementioned transformation in a lower dimension is defined  as:\\
\textbf{Problem 2:} 
\begin{equation}\label{eqn:Edeepc_general}
\begin{aligned}
    \min_{\substack{\bar{g},\sigma\\u_{pred},y_{pred}}} \; & f_1(u_{pred},y_{pred},\sigma,\bar{g}) \\
    \text{s.t.}\; &\bar{H}\bar{g}=\begin{bmatrix}
        y_{init}\\y_{pred}\\u_{init}\\u_{pred}
    \end{bmatrix}+\sigma, \\
    & f_3(y_{pred}, u_{pred}, \sigma) \leq 0
\end{aligned}
\end{equation}
where $\bar{H} := U_1 \Sigma$, signifying the transformed version of the Hankel matrix.

\begin{lemma} \label{lemma:deepc_equal}
Problem 1 and Problem 2 are equivalent. The optimal values of $y_{pred}, u_{pred}$ for Problem 1 are also optimal for Problem 2, and vice versa.
\end{lemma} 

\begin{proof}
Problem 1  can change the decision variable $g$ by:
\begin{align*}
    \tilde{g} = \begin{bmatrix}\tilde{g}_1 \\ \tilde{g}_2 \end{bmatrix} =  
    \begin{bmatrix}V_1^{\top} g \\ V_2^{\top} g \end{bmatrix} = \begin{bmatrix}V_1 & V_2\end{bmatrix}^{\top}g
\end{align*}
because $\begin{bmatrix}V_1 & V_2\end{bmatrix}$ is an orthogonal matrix~\cite{boyd2004convex}.
Then because $Hg = U_1 \Sigma V_1^{\top}g = \bar{H}\tilde{g}_1 $, the objects and constraints in the new equivalent problem are separable with respect to $\tilde{g}_1$ and $\tilde{g}_2$:
\begin{align*}
    \min_{\substack{\tilde{g},\sigma\\u_{pred},y_{pred}}} \; & f_1(u_{pred},y_{pred},\sigma,\tilde{g}_1) +  f_2(\tilde{g}_2) \\
    \text{s.t.}\; & \bar{H}\tilde{g}_1=\begin{bmatrix}
        y_{init}\\y_{pred}\\u_{init}\\u_{pred}
    \end{bmatrix}+\sigma, \\
    & f_3(y_{pred}, u_{pred}, \sigma) \leq 0
\end{align*}
Therefore, we can solve them separately by:
\begin{align*}
    \min_{\substack{\tilde{g}_1,\sigma\\u_{pred},y_{pred}}} \; & f_1(u_{pred},y_{pred},\sigma,\tilde{g}_1)\\
    \text{s.t.}\; &\bar{H}\tilde{g}_1=\begin{bmatrix}
        y_{init}\\y_{pred}\\u_{init}\\u_{pred}
    \end{bmatrix}+\sigma, \\
    & f_3(y_{pred}, u_{pred}, \sigma) \leq 0 \\
    \min_{\tilde{g}_2} \; &  f_2(\tilde{g}_2)
\end{align*}
By replacing $\tilde{g}_1$ by $\bar{g}$ in the first sub-problem above, we get Problem 2.
\end{proof}

Leveraging Lemma~\ref{lemma:deepc_equal}, we can deduce the low-dimensional version of the L2-DeePC~\eqref{eqn:deepc_L2}. This inference is drawn from the relationship:
$\lVert g\rVert_2^2 = g^{\top}\begin{bmatrix}V_1 & V_2\end{bmatrix} \begin{bmatrix}V_1 & V_2\end{bmatrix}^{\top}g = \lVert V_1 g\rVert_2^2 + \lVert V_2 g\rVert_2^2$:
\begin{equation}\label{eqn:Edeepc_L2}
\begin{aligned}
         \min_{\substack{\bar{g},\sigma\\u_{pred},y_{pred}}} \; & J(u_{pred},y_{pred}) + \lambda_{\sigma}\lVert\sigma\rVert_2^2 + \lambda_g \lVert \bar{g}\rVert_2^2 \\
        \text{s.t.}\; &\bar{H}\bar{g}=\begin{bmatrix}
        y_{init}+\sigma\\y_{pred}\\u_{init}\\u_{pred}
    \end{bmatrix}, \\
    & y_{pred} \in \mathbb{Y}, u_{pred} \in \mathbb{U}
\end{aligned}
\end{equation}

\begin{remark}
In Problem 2, the decision variable $\bar{g}$, which belongs to $\mathbb{R}_{r_H}$, is independent of the columns of the Hankel matrix.
\change{The authors in~\cite{chen2022data,LOWDEEPCzhang2022dimension} introduce the same SVD-based transformation and establish the equivalence using KKT conditions~\cite{LOWDEEPCzhang2022dimension}. However, their study is limited to L2-DeePC, which is a special case of the more general form of DeePC~\eqref{eqn:deepc_general} described in our study.
}.

\end{remark}



\subsection{Efficient SVD updates}
In the preceding section, we established that the general-form DeePC~\eqref{eqn:deepc_general} can be converted into a more compact, low-dimensional format~\eqref{eqn:Edeepc_general} via SVD. Notably, the dimensionality of the decision variable in~\eqref{eqn:Edeepc_general} is governed solely by the rank of the Hankel matrix $H$.

Expanding upon this, the current section introduces a rapid SVD updating technique~\cite{brand2006fast,bunch1978updating}. This method obviates the need for a complete SVD recalculation with each recursive update~\eqref{eqn:recurive_update}.
When the previous SVD components, specifically $U_{1}$ and $\Sigma$, are available and $H$ undergoes an update as per~\eqref{eqn:recurive_update}, Algorithm~\ref{alg:SVD} can be leveraged to update the new $U_{1}$ and $\Sigma$.



\begin{algorithm} 
  \caption{Fast SVD updating}
  \label{alg:SVD} 
{
\textbf{Given:} Current SVD components: $U_1, \Sigma$
\begin{itemize} [nolistsep,leftmargin=1.5em]
    \item[1)] Retrieve \change{the column $a$ to be added to} $H$, i.e. $\begin{bmatrix}
        y_{t-L:t-1}\\ u_{t-L:t-1}
        \end{bmatrix}$ at time $t$. Compute $r=\text{rank}(\Sigma)$. 
    \item[2)] If $r<\textit{row}_{H}$, update $U_1, \; \Sigma$ by~\cite{brand2006fast}: 
    \begin{itemize} [nolistsep,leftmargin=1.0em]
        \item[] Compute $m = U^{\top}a, \; p = a - Um, \; R_a = \lVert p \lVert_2, \; P = R_a^{-1}p$
        \item[] Compute
        \resizebox{0.2\hsize}{!}{%
        $K = \begin{bmatrix}  \Sigma & m \\ 0 & R_a  \end{bmatrix}$%
        }
        and its SVD: $K = C \bar{\Sigma} D^{\top}$ 
        \item[] If $\text{rank}(\bar{\Sigma})==r$: 
        \begin{itemize} [nolistsep,leftmargin=1.0em]
            \item[] $U_1 \leftarrow \begin{bmatrix}U_1 & P \end{bmatrix}C(:,1:r), \;   \Sigma \leftarrow \bar{\Sigma}(1:r,1:r)$
        \end{itemize}
        \item[] If $\text{rank}(\bar{\Sigma})==r+1$: 
        \begin{itemize} [nolistsep,leftmargin=1.0em]
            \item[] $U_1 \leftarrow \begin{bmatrix}U_1 & P \end{bmatrix}C, \;  \Sigma \leftarrow \bar{\Sigma}$
        \end{itemize}   
    \end{itemize}
    \item[3)] If $r==\textit{row}_{H}$, update $U_1, \; \Sigma$ by~\cite{bunch1978updating}:
    \begin{itemize} [nolistsep,leftmargin=1.0em]
        \item[] Compute $z = U_1^{\top}a$
        \item[] Compute eigendecomposition of $\Sigma^{2} + zz^{\top}$: $C\bar{\Sigma}C^{\top}$
        \item[] Update $U_1 \leftarrow U_1C$, $\Sigma \leftarrow \sqrt{\bar{\Sigma}}$
    \end{itemize} 
\end{itemize}
}
\end{algorithm}

\begin{lemma} \label{lem:svd_load}
\change{
    Algorithm~\ref{alg:SVD} calculates $U_1$ and $\Sigma$ identical to the results obtained through the direct SVD of $H$ following each recursive update~\eqref{eqn:recurive_update}.  It boasts a computational complexity of $\mathcal{O}(\textit{row}_{H}r_{H}^2)$ and a space requirement of $\mathcal{O}(\textit{row}_{H}r_{H})$. 
}
\end{lemma}
\begin{proof}
\change{
    Proofs of equivalence for the conditions in steps $2)$ and $3)$ of Algorithm~\ref{alg:SVD} are provided in~\cite{brand2006fast,bunch1978updating}, which is omitted due to the limited space. All the matrix multiplications requiries a load of $\mathcal{O}(\textit{row}_{H}r_{H}^2)$. In addition, the SVD's load in step $2)$ is $\mathcal{O}(r_{H}^3)$ (or $\mathcal{O}(r_{H}^2)$ due to the special structure~\cite{brand2006fast}), and the eigendecomposition's load is $\mathcal{O}(\textit{row}_{H}^3)$. Because $r_{H} < \textit{row}_{H}$ for step $2)$ and $r_{H} = \textit{row}_{H}$ for step $3)$, the overall complexity is $\mathcal{O}(\textit{row}_{H}r_{H}^2)$. Finally, the matrices' size directly determines the space requirement. 
    }
\end{proof}

\subsection{Conclusion of the algorithm}

A computationally efficient recursive DeePC is summarized in Algorithm~\ref{alg:ERdeepc}. The general form of DeePC represented in~\eqref{eqn:deepc_general} undergoes a transformation into a low-dimensional equivalent as outlined in~\eqref{eqn:Edeepc_general}, using SVD. Moreover, with each successive update as indicated in~\eqref{eqn:recurive_update}, the new SVD components are rapidly updated. 

\begin{algorithm} 
  \caption{Efficient Recursive DeePC}
  \label{alg:ERdeepc} 
{
\begin{itemize} [nolistsep,leftmargin=1.5em]
    \item[0)] Retrieve some persistently excited past I/O data. Construct $H$ and compute its SVD. Build the initial low-dimensional DeePC controller based on the Problem 2, such as~\eqref{eqn:Edeepc_L2}.
    \item[1)] Retrieve the recent $L$-step measurements. Update the SVD components based on Algorithm~\ref{alg:SVD} and update $\bar{H}$.
    \item[2)] Retrieve the recent $t_{init}$-step measurements. Solve the low-dimensional DeePC and apply the optimal input as $u_t=u_{pred}^{\ast}(1)$.
    \item[3)] Pause until the subsequent sampling time, update $t\leftarrow t+1$ and revert to step 1.
\end{itemize}
}
\end{algorithm}

\begin{lemma} \label{lemma:alg_equal}
If the DeePC in the form of Problem 1 is used in Algorithm~\ref{alg:Rdeepc},
\change{
then Algorithm~\ref{alg:Rdeepc} and Algorithm~\ref{alg:ERdeepc} are equivalent. 
}
\end{lemma} 
\begin{proof}
\change{
    At the initial step, Problem 1 and Problem 2 are respectively constructed in two Algorithms, which have been proved to be equivalent in Lemma~\ref{lemma:deepc_equal}. After the first recursive update~\eqref{eqn:recurive_update}, the new SVD components are exactly updated by Algorithm~\ref{alg:SVD} proved in Lemma~\ref{lem:svd_load}. Therefore, Problem 1 and Problem 2 are still equivalent by Lemma~\ref{lemma:deepc_equal}. The proof is then completed by induction.
}
\end{proof}

As a result, the total computational complexity depends mainly on $\textit{row}_{H}$ and $r_{H}$. \change{Because $\textit{row}_{H} = (n_{init}+n_{pred})(n_u+n_y)$, and
the decision variables $\bar{g} \in \mathbb{R}_{r_H}$, $u_{pred} \in \mathbb{R}_{n_u(n_{init}+n_{pred})}$, $y_{pred} \in \mathbb{R}_{n_y(n_{init}+n_{pred})}$, $\sigma \in \mathbb{R}_{\textit{row}_{H}}$ in~Problem 1. Besides, the complexity of the fast SVD updating method is proved in Lemma~\ref{lem:svd_load}.
Therefore, the complexity is fixed after the DeePC's parameter is settled as $r_{H} \leq \textit{row}_{H}$. 
}
It's notable that the size of the original recursive DeePC in Algorithm~\ref{alg:Rdeepc} relates to \change{$\textit{col}_{H} = T-\textit{row}_{H}+1$}, which increases with the addition of more data.

The computational burden of Algorithm~\ref{alg:ERdeepc} is comparable to the recursive SPC method~\cite{CLSPCdong2008closed}. The latter has a decision variable in its sparse representation of size $\textit{row}_H$, which can recursive update using \textit{Recursive Least Square} at a computational complexity of $\mathcal{O}(\textit{row}_H^2)$\cite{verheijen2022recursive}. \change{In the next Section, we will prove that SPC is equivalent to a specialized DeePC belonging to the general-from DeePC~\eqref{eqn:deepc_general}, adaptable as well by Algorithm~\ref{alg:ERdeepc}.}

\begin{remark}
Algorithm~\ref{alg:ERdeepc} offers extensions to other adaptive DeePC strategies, typified by references like~\cite{lian2021adaptive,berberich2022linear}. These strategies are applicable to slowly time-varying linear systems or approximate dynamics of unknown nonlinear systems across varied operating points. Other fast SVD modifications, such as the integration of forgetting factors and downdating~\cite{gu1995downdating}, can be incorporated for extensions to these adaptive methods (see Appendix~\ref{apenendix:C}).
\end{remark}

\jc{low rank, full rank\\ 
downdates, forgetting factor\\
algorithm}

\jc{remark: apply to other DeePC, SPC\\
comparison of RLS}

\section{\change{An extension to data-driven methods based-on Pseudoinverse}}\label{sect:4}
\jc{1 deepc; 1 low-dimen version; Lemma: open-loop (SPC real dynamics), closed-loop (in Appendix: based-on the paper closed-loop SPC and previous work ) }

A pivotal strength of Algorithm~\ref{alg:ERdeepc} is its versatile nature, rooted in its generic DeePC formulation. Beyond encompassing various DeePC variants~\cite{coulson2019regularized,huang2019data,lian2021adaptive}, it holds potential for extension to various data-driven methodologies that utilize the Hankel matrix, such as simulation~\cite{markovsky2008data}, physics-based filters~\cite{lian2023physically} and
data-driven observers~\cite{turan2021data,shi2022data}. 

Among them, A group of researchers utilizes Pseudoinverse to achieve prediction or estimation~\cite{huang2019data,markovsky2008data,turan2021data,shi2022data}.
For the purpose of this section, we will illustrate that these Pseudoinverse-based methods can be generalized in the form of Problem 1 using a specific data-driven prediction formulation~\cite{huang2019data}. Next, we will demonstrate that Problem 1 can generalize SPC, which also employs Pseudoinverse and Hankel matrices. Additionally, we present how to achieve asymptotically consistent prediction for stochastic LTI systems through recursive data updates using Algorithm~\ref{alg:ERdeepc}.


\subsection{Comparison to Data-driven prediction}

Given measured $u_{init},  y_{init}$ and required $u_{pred}$, a data-driven prediction method based on Pseudoinverse~\cite{huang2019data} is formulated as:
\begin{equation} \label{eqn:pse_deepc}
\begin{aligned}
   y_{pred} &= H_{y,pred}g\\
    g &= \begin{bmatrix}
        H_{y,init}\\H_{u}
    \end{bmatrix}^{\dag}\begin{bmatrix}
        y_{init}\\u_{init}\\u_{pred}
    \end{bmatrix} 
\end{aligned}    
\end{equation}
where the sub-Hankel matrices are derived from the original Hankel matrix by
$H_y = \begin{bmatrix}
        H_{y,init}\\H_{y,pred}
    \end{bmatrix}$.    
The matrix $H_{y,init}$ is of depth $n_{init}$ and the depth of $H_{y,pred}$ is the prediction horizon $n_{pred}$ such that $n_{init}+n_{pred} = L$. 

Consider an optimization problem:
\begin{equation} \label{eqn:bilevel_prediction}
\begin{aligned}
     (y_{pred},g) = &\argmin_{y_{pred,l},g_l} \lVert g_l \rVert_2^2 \\
    & \text{s.t.}\; H g_l = \begin{bmatrix}
        y_{init}\\y_{pred,l}\\u_{init}\\u_{pred}
    \end{bmatrix}
\end{aligned}
\end{equation}
\changeE{
\begin{assumption} \label{assmp:pred_feasible}
    The problem~\eqref{eqn:bilevel_prediction} is feasible.
\end{assumption}
}

\begin{lemma}
    Under Assumption~\ref{assmp:pred_feasible}, \eqref{eqn:bilevel_prediction} computes the same unique optimal solution as~\eqref{eqn:pse_deepc}.
\end{lemma}
\begin{proof}
Consider a problem:
    \begin{align*}
        y_{pred} &= H_{y,pred}g   \\
     g &=\argmin_{g_l} \lVert g_l \lVert^2 \\
    & \quad \text{s.t.}\; \begin{bmatrix}
        H_{y,init}\\H_{u}
    \end{bmatrix}g_l = \begin{bmatrix}
        y_{init}\\u_{init}\\u_{pred}
    \end{bmatrix}
    \end{align*}    
    
\noindent Under Assumption~\ref{assmp:pred_feasible}, the problem remains feasible. Given that the Pseudoinverse offers a unique solution with the minimal L2-norm for least squares problems~\cite{penrose1956best}, this problem yields the same prediction as~\eqref{eqn:pse_deepc}. The proof concludes with the observation that this problem and~\eqref{eqn:bilevel_prediction} can be equivalently expressed by the same linear equations when applying the KKT conditions.   
\end{proof}

By the fact $\lVert g\rVert_2^2 = g^{\top}\begin{bmatrix}V_1 & V_2\end{bmatrix} \begin{bmatrix}V_1 & V_2\end{bmatrix}^{\top}g = \lVert V_1 g\rVert_2^2 + \lVert V_2 g\rVert_2^2$ and adding an equality constraint so that $u_{pred}$ is equal to the required value, we can write~\eqref{eqn:bilevel_prediction} in the form of Problem 1.
\changeE{
\begin{remark}  When the input sequence for $H_u$ is persistently excited, Assumption~\ref{assmp:pred_feasible} is considered mild. Because then the assumption is satisfied in noiseless conditions as confirmed by Lemma~\ref{lemma:funda}. In the presence of noise, this assumption also holds  since Hankel matrices are almost certainly of full row rank.
    Alternatively, Appendix~\ref{apenendix:D} presents some optimization formulations and their low-dimensional equivalents, which are always feasible and provide the same solution as~\eqref{eqn:pse_deepc}. For other Pseudoinverse-based data-driven methods~\cite{markovsky2008data,turan2021data,shi2022data}, similar results can be derived after little modification of~\eqref{eqn:pse_deepc}.
\end{remark}}

\subsection{Comparison to SPC} \label{sect:4B}


This section describes how to generalize SPC in the form of Problem 1. In addition, Algorithm~\ref{alg:ERdeepc} helps to achieve asymptotically consistent prediction by continuously involving open-loop and closed-loop data online. The SPC controller~\cite{SPCfavoreel1999spc} is formulated as,
\begin{subequations} \label{eqn:spc}
\begin{align}
\min_{u_{pred},y_{pred}} \; & J(u_{pred},y_{pred}) \nonumber\\
\text{s.t.}\; & y_{pred} \in \mathbb{Y}, u_{pred} \in \mathbb{U} \nonumber \\
    & y_{pred} = K\begin{bmatrix}
            y_{init}\\u_{init}\\u_{pred}
        \end{bmatrix}  \label{eqn:spc_1} \\
    & K = H_{y,pred}\begin{bmatrix}
        H_{y,init}\\H_{u}
    \end{bmatrix}^{\dag} \label{eqn:spc_2}
\end{align}
\end{subequations}

Based on the specific-form data-driven prediction~\eqref{eqn:bilevel_prediction}, a bi-level DeePC is defined as:
\begin{equation} \label{eqn:bilevel_deepc_op}
\begin{aligned}
\min_{\substack{g,u_{pred}\\y_{pred}}} \; & J(u_{pred},y_{pred}) \\
\text{s.t.}\; & y_{pred} \in \mathbb{Y}, u_{pred} \in \mathbb{U}  \\
    & \eqref{eqn:bilevel_prediction}
\end{aligned}
\end{equation}

\begin{lemma} \label{lemma:deepc_spc3}
Under Assumption~\ref{assmp:pred_feasible}, SPC~\eqref{eqn:spc} and the DeePC~\eqref{eqn:bilevel_deepc_op} are equivalent.
\end{lemma}
\begin{proof}
    The only difference between the two controllers is their prediction parts, i.e.~\eqref{eqn:spc_1} and~\eqref{eqn:spc_2}, \eqref{eqn:bilevel_prediction}. The fact that both predictions can be written in the same explicit form finishes the proof: $  y_{pred} = H_{y,pred} \begin{bmatrix}
        H_{y,init}\\H_{u}
    \end{bmatrix}^{\dag}\begin{bmatrix}
        y_{init}\\u_{init}\\u_{pred}
    \end{bmatrix}$
\end{proof}
\changeE{
\begin{remark}  
A bi-level DeePC that is equivalent to the SPC~\eqref{eqn:spc}, without requiring any assumption, is detailed in the Appendix~\ref{apenendix:E} due to limited space. In addition, some low-dimensional equivalents and an efficient recursive SPC algorithm are introduced.
\end{remark}}

The analysis of consistent prediction is as follows. Consider a stochastic LTI system defined in innovation form:
\begin{equation}
\begin{aligned} \label{eqn:SLTI}
    x_{t+1} = Ax_t + Bu_t + Ke_t \\
    y_t = Cx_t + Du_t + e_t
\end{aligned}    
\end{equation}
where $K$ denotes the Kalman gain and $e_k$ is a zero-mean white noise signal.
The prediction $y_{pred}$ at time $t$ is consistent if its expectation is an unbiased estimation of the real output sequence $y_{real}$~\cite{CLDEEPCdinkla2023closed,SIDvan2012subspace}, i.e.
\begin{align*}
    \mathbb{E}_{e} (y_{pred}-y_{real}) = \mathbf{0}
\end{align*}

\begin{assumption}\label{assmp:Ak_stable}
The Kalman gain K in~\eqref{eqn:SLTI} ensures that the matrix $A-KC$ is strictly stable. The initial step $n_{init}$ is sufficiently large so that $(A-KC)^{t_{init}} \approx 0$.
\end{assumption}

\begin{assumption}\label{assmp:input_infnity}
The input is quasi-stationary
so that limits of time averages of the input
sequence exists.
\end{assumption}

\begin{assumption}\label{assmp:H_full_rank}
The input sequence $\{u_i\}_{i=1}^T$ for Hankel matric $H_u$ is persistently exciting of order $L+n_x$.
\end{assumption}

\jc{write prediction and deepc separately; \\
prediction, lemma\\
open-loop deepc\\
closed-loop deepc\\
full rank: remark, regularization}

\jc{definition: consistent prediction}

\begin{lemma} \label{lemma:deepc_spc1}
    Under Assumptions~\ref{assmp:Ak_stable}, \ref{assmp:input_infnity} and~\ref{assmp:H_full_rank}, \eqref{eqn:bilevel_prediction} constructed by open-loop data provides a consistent prediction when $\textit{col}_H \rightarrow \infty$.
\end{lemma}

\begin{lemma} \label{lemma:deepc_spc2}
    Assumptions~\ref{assmp:Ak_stable}, \ref{assmp:input_infnity} and~\ref{assmp:H_full_rank} stand. Assume that $D=0$ in the LTI system or the I/O data is collected by feedback control with at least one sample time delay. Then \eqref{eqn:bilevel_prediction} with $n_{pred}=1$ constructed by the closed-loop data provides a consistent prediction when $\textit{col}_H \rightarrow \infty$
\end{lemma}



The proof of Lemmas~\ref{lemma:deepc_spc1} and~\ref{lemma:deepc_spc2} is elaborated in Appendix~\ref{apenendix:lemma_proof}.
Assumptions~\ref{assmp:Ak_stable}, \ref{assmp:input_infnity} and~\ref{assmp:H_full_rank} used therein also frequently emerge in consistency analysis in the field of system identification, as seen in~\cite{SIDvan2012subspace,SPCfavoreel1999spc}. 
In addition,  because Lemma~\ref{lemma:deepc_spc2} guarantees asymptotically consistent prediction for~\eqref{eqn:bilevel_prediction}  with closed-loop data when setting $n_{pred}=1$, one can successively apply~\eqref{eqn:bilevel_prediction} with $n_{pred}=1$ to achieve consistent multi-step output prediction via:

\begin{equation}\label{eqn:bilevel_prediction_cl}
\begin{aligned}
    & \forall i = 1,2,\dots,n_{pred}:\\
    & \quad  (y_{pred}(i), g_i)=\argmin_{y_{pred,l},g_l} \lVert g_l \rVert_2^2 \\
    & \quad \quad \text{s.t.} \; H g = \begin{bmatrix}
        y_{init}(i:n_{pred})\\y_{pred}(1:i-1)\\ y_{pred,l}\\u_{init}(i:n_{pred})\\u_{pred}(1:i)
    \end{bmatrix} 
\end{aligned}
\end{equation}
where $y_{pred}(i)$ represents the $i-$th output in $y_{pred}$, and $y_{pred}(i:j)$ captures the vector from the $i$-th to $j-$th outputs within $y_{pred}$ (with similar notations applied elsewhere). Similar setups have been validated in prior DeePC and SPC studies~\cite{o2022data,CLSPCdong2008closed}.
A new bi-level DeePC can be defined as:
\begin{equation} \label{eqn:bilevel_deepc_cl}
\begin{aligned}
\min_{\substack{g,u_{pred}\\y_{pred}}} \; & J(u_{pred},y_{pred}) \\
\text{s.t.}\; & y_{pred} \in \mathbb{Y}, u_{pred} \in \mathbb{U}  \\
    & \eqref{eqn:bilevel_prediction_cl}
\end{aligned}
\end{equation}

Notably, one can directly apply Algorithm~\ref{alg:ERdeepc} to recursively update the bi-level DeePC~\eqref{eqn:bilevel_deepc_cl}. A tractable computation method for this bi-level DeePC is to transform the lower-level problem into linear constraints by KKT conditions. More details can be referred to our previous work~\cite{lian2021adaptive}. According to Lemma~\ref{lemma:deepc_spc2}, with an infinite length of closed-loop data, it's feasible to obtain an unbiased output prediction for the stochastic LTI system. Nonetheless, it's important to note that there may not be a monotonic improvement in prediction and control performance throughout the update cycle.

In addition, one can adapt Algorithm~\ref{alg:ERdeepc} to update the data-driven prediction online in the two bi-level DeePCs~\eqref{eqn:bilevel_deepc_op} and~\eqref{eqn:bilevel_deepc_cl} with data from other controllers. To achieve this, the modification required for~\eqref{eqn:bilevel_prediction_cl} (or~\eqref{eqn:bilevel_prediction}) is simply replacing the DeePC in step 3) of Algorithm~\ref{alg:SVD} with other closed-loop controllers (or open-loop control signals).



\jc{not necessary better then open-loop in the process; error may be extended due to recursive model; could use recent data to do prediction test to check if it's better; after some time use this method}


\section{Simulation}\label{sect:5}

In this section, we evaluate the effectiveness of the proposed efficient recursive DeePC methodology through simulation studies.
We utilize a discrete-time LTI system, as detailed in~\cite{SPCfavoreel1999spc}, which models two circular plates coupled with flexible shafts. The system's matrices, conforming to~\eqref{eqn:SLTI}, are provided:
\begin{equation*}
    \resizebox{1.0\hsize}{!}{%
    $ \begin{aligned}
    &A = \begin{bmatrix}
        4.4 & 1 & 0 & 0 & 0\\
        -8.09 & 0 & 1 & 0 & 0\\
        7.83 & 0 & 0 & 1 & 0\\
        -4 & 0 & 0 & 0 & 1\\
        0.86 & 0 & 0 & 0 & 0
    \end{bmatrix}, \; 
    B= \begin{bmatrix}0.00098\\0.01299\\0.01859\\0.0033\\-0.00002 \end{bmatrix}, K = \begin{bmatrix}2.3\\ -6.64\\ 7.515\\ -4.0146 \\0.86336 \end{bmatrix}\\
    &C = \begin{bmatrix}1& 0& 0& 0& 0 \end{bmatrix}, \;D = 0
    \end{aligned}
    $%
        }
\end{equation*}
During the simulation, the noise variance is set to $\text{var}(e_t) = 0.1$, and the input is restricted to $|u_t| \leq 10$. The optimization problems in the following simulation are solved by the solver \textsc{quadprog}  in \textsc{MATLAB} with Intel Core i7-1165G7 2.80 GHz processor.

\subsection{Validation of Algorithm~\ref{alg:ERdeepc}}

To initiate, a 200-step trajectory is generated with the input defined as a zero-mean white noise signal, having $\text{var}(u_t) = 1$. Employing this initial trajectory, a L2-DeePC~\eqref{eqn:deepc_L2} is established, targeting the objective $J(u_{pred},y_{pred})= \lVert y_{pred}-ref\rVert_2^2 + 0.001\lVert u_{pred}\rVert_2^2$. Initially, the reference is set at $10$ for 1000 steps and subsequently adjusted to $0$ for the ensuing 1000 steps. The parameters are designated as $\lambda_{\sigma} = 10^6$, $\lambda_{g} = 10^4$, and $n_{init} = n_{pred} = 10$. These parameters aren't meticulously tuned, as our primary interest lies in evaluating the efficiency of Algorithm~\ref{alg:ERdeepc}. 
The L2-DeePC controls the system and is recursively updated by Algorithm~\ref{alg:Rdeepc}.
For comparative analysis, an identical procedure is employed utilizing Algorithm~\ref{alg:ERdeepc}, integrated with the equivalent low-dimensional L2-DeePC, as delineated in~\eqref{eqn:Edeepc_L2}.

\begin{figure}[!ht]
    \centering
    \includegraphics[width=1.0\linewidth]{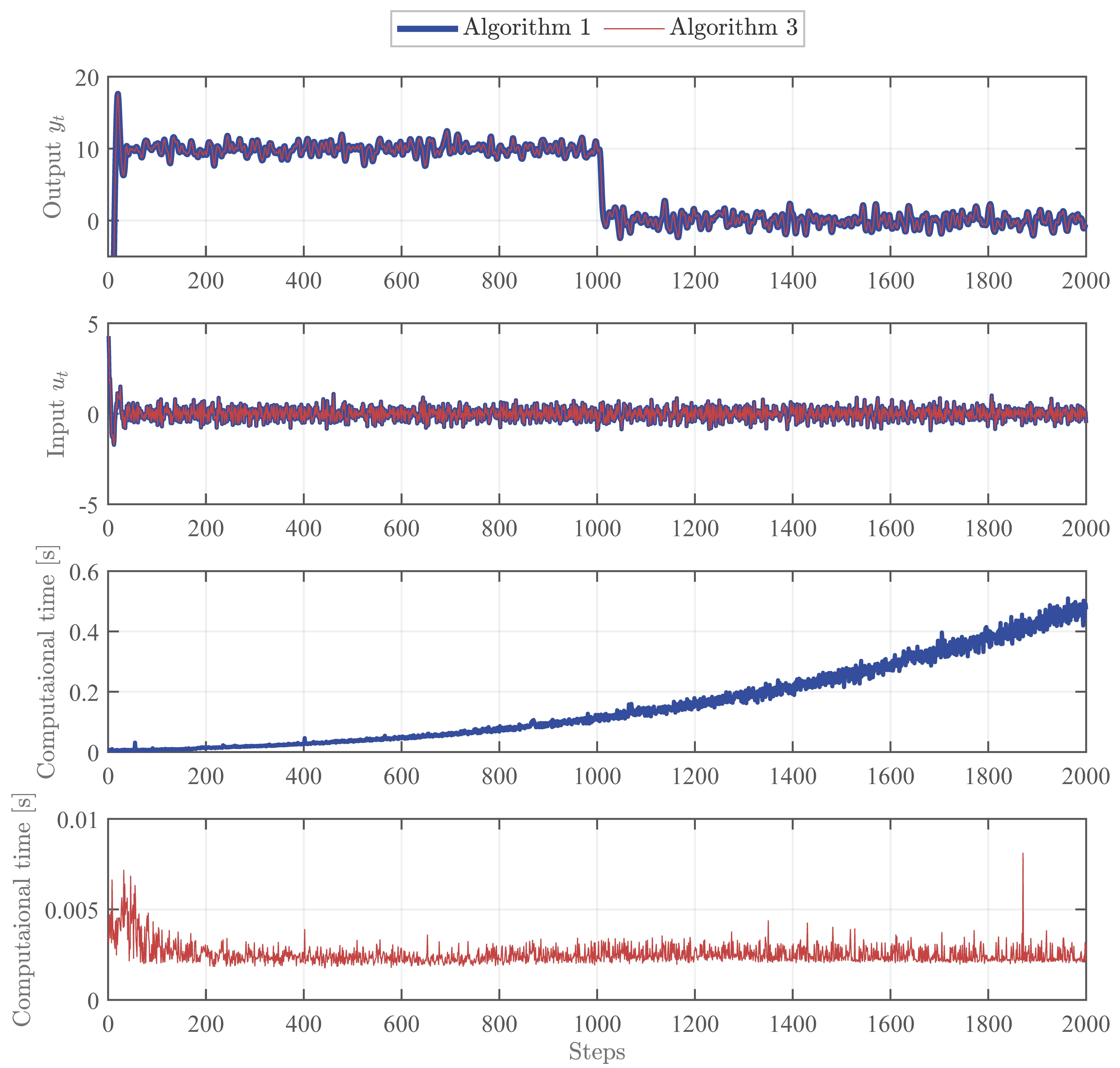}
    \caption{Comparison of Algorithm 1 and Algorithm 3.}
    \label{fig:equal}
\end{figure}

Table~\ref{tab:simulation_1} provides the statistical results from 10 Monte Carlo simulations across different noise scenarios. Both algorithms yield almost identical input and output signals, with only slight numerical errors, Algorithm~\ref{alg:ERdeepc} proves to be faster in execution than Algorithm~\ref{alg:Rdeepc}.
For a closer look, Figure~\ref{fig:equal} shows the resulting trajectories for a specific noise scenario.
The input and output trajectories validate the equivalence between the two algorithms. Additionally, we analyze the computational time required for each recursive update and optimization for both algorithms. We notice that the time for Algorithm~\ref{alg:Rdeepc} increases superlinearly as more data are added, whereas the time for Algorithm~\ref{alg:ERdeepc} stays relatively steady, highlighting its efficiency.


\vspace{0.5em}

\begin{table}[!ht]
\renewcommand{\arraystretch}{1.3}
\centering 
\caption{Statistical results of 10 Monte Carlo runs: the differences of input ($e_u = \lvert u_{1} - u_{3}  \rvert$) and output ($e_y = \lvert y_{1} - y_{3}  \rvert)$), the computational time of each recursive step ($time$), where $\cdot_{1}$ and $\cdot_{3}$ respectively indicate the data from Algorithm 1 and 3.}  
\label{tab:simulation_1}
\begin{tabular}{  l  l  l l}
  \hline
  Average $e_u$ & Average $e_y$ & Average $time_1$ & Average $time_3$ \\ 
  \hline
 $6.7\times10^{-12}$ & $5.2\times10^{-12}$ & 0.1557 [s] & 0.0026 [s] \\ 
  \hline  
\end{tabular}
\end{table}

\subsection{Comparison to SPC}

This section evaluates the asymptotic consistency of the data-driven prediction in equations~\eqref{eqn:bilevel_deepc_op} and~\eqref{eqn:bilevel_deepc_cl} by using Algorithm~\ref{alg:ERdeepc}. We will refer them to as DDP1 and DDP2 for brevity. The equivalence between~\eqref{eqn:bilevel_deepc_op} and SPC~\eqref{eqn:spc} is also tested. 
Given that all the data-driven prediction methods and the ground truth (see Appendix~\ref{apenendix:B}) can be expressed in the matrix form:
\begin{align} \label{eqn:ypred_spec}
    y_{pred} = K_{y,init}y_{init} + K_{u,init}u_{init} + K_{u,pred}u_{pred}
\end{align}
, consistency is tested by comparing discrepancies among the involved matrices. In this study, we set $n_{init} = n_{pred} = 50$, with a large $n_{init}$ ensuring compliance with Assumption~\ref{assmp:Ak_stable}

\begin{figure}[!ht]
    \centering
    \includegraphics[width=1.0\linewidth]{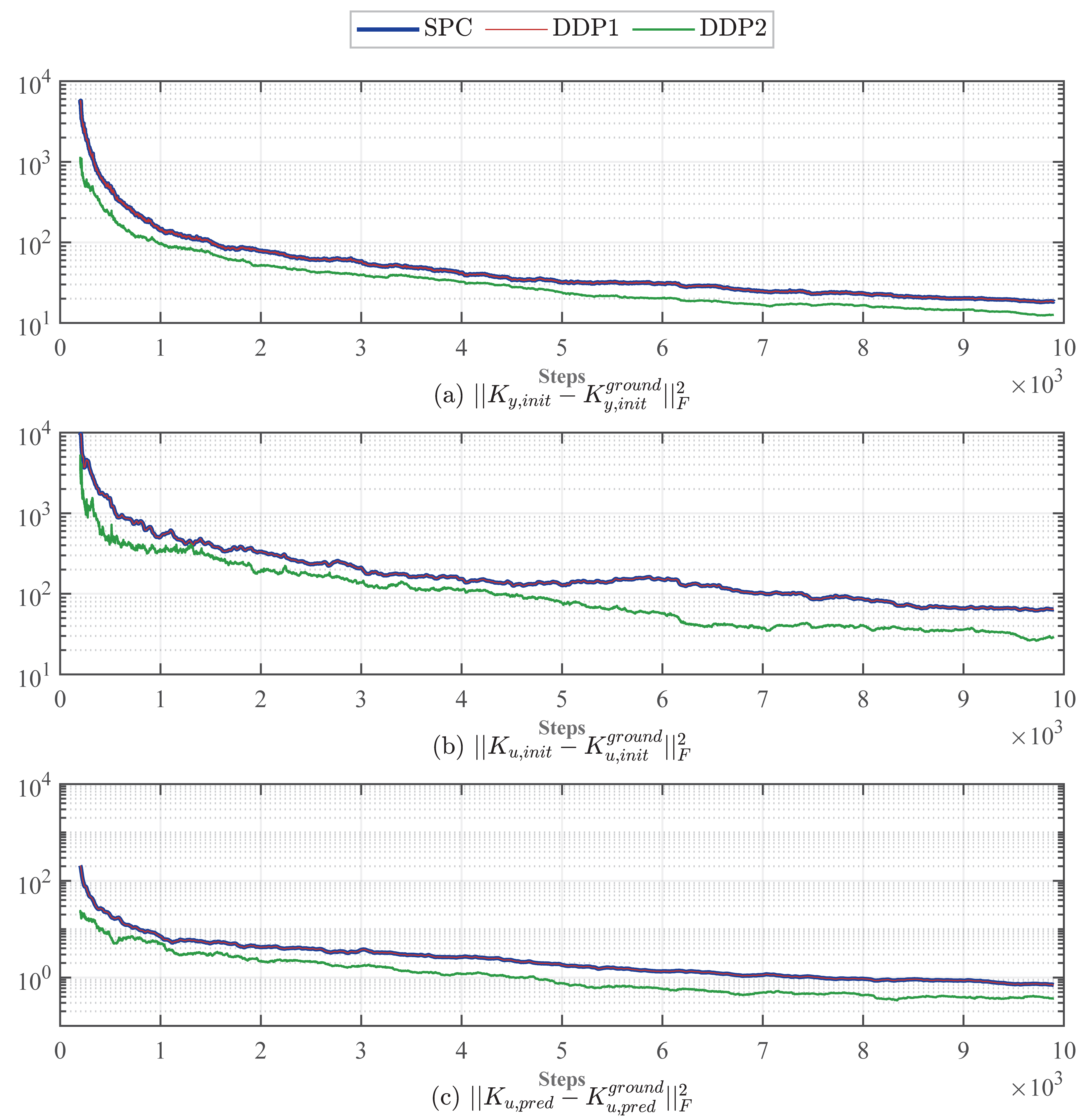}
    \caption{Consisteny analysis by open-loop data. SPC: from~\eqref{eqn:spc_1} and~\eqref{eqn:spc_2}; DDP1: from~\eqref{eqn:bilevel_deepc_op}; DDP2: from~\eqref{eqn:bilevel_deepc_cl}. $K_{\cdot}^{\text{ground}}$
    indicates the matrix from the ground truth.
     }
    \label{fig:open}
\end{figure}

Firstly, an open-loop trajectory spanning 10000 steps is generated using an input characterized as a zero-mean white noise signal with a variance $\text{var}(u_t) = 1$. DPP1, DDP2, and SPC are initialized using 150 steps to assemble the Hankel matrix. They are then efficiently updated in a recursive manner, leveraging a variant of Algorithm~\ref{alg:ERdeepc} as outlined at the end of Section~\ref{sect:4B}.
Figure~\ref{fig:open} depicts average outcomes from 10 Monte Carlo simulations, wherein the deviation from the ground truth is calculated at each iteration. As more open-loop data are incorporated into the three prediction methods, the matrix discrepancies consistently diminish, reinforcing their validity. Furthermore, the equivalence between SPC and DPP1 is validated.

\begin{figure}[!ht]
    \centering
    \includegraphics[width=1.0\linewidth]{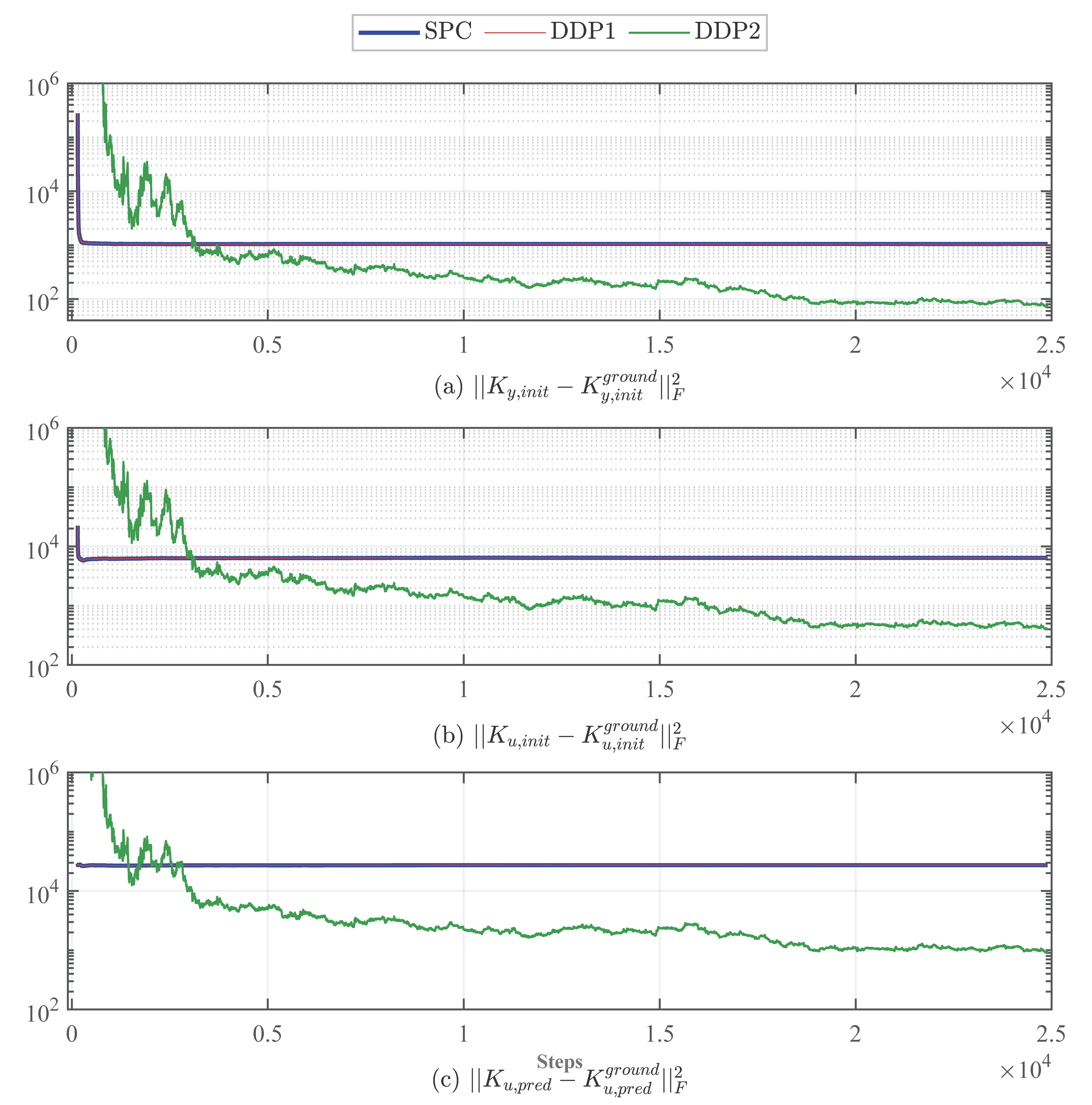}
    \caption{Consisteny analysis by closed-loop data}
    \label{fig:closed}
\end{figure}

The subsequent experiment employs a 25000-step closed-loop trajectory, controlled by a static DeePC constructed from the open-loop trajectory of the previous test. Average results over 10 random noise signal realisations are showcased in Figure~\ref{fig:closed}. The matrix discrepancies from the ground truth, as observed in the SPC and DDP1, initially decline but later stabilize. Conversely, DDP2 continually exhibits a reduction in matrix differences as it integrates more closed-loop data. However, it's notable that when the Hankel matrix lacks sufficient data, matrix discrepancies in DDP2 exceed others', and its improvement rate lags behind that observed in Figure~\ref{fig:open}. Future work will focus on optimizing the closed-loop controller design to expedite improvements in DDP2.

\jc{Simulation: open-loop 1w steps, closed-loop 10w steps \\
1. Equal performance of 2 recursive DeePC, SVD comparison by Matlab and ours (initialized by open-loop data, then updated by closed-loop data)\\
2. Use the closed-loop data to do a prediction check \\
3. Use common R-DeePC and the CP-DeePC to run new performance
}

\section{Conclusion}
In conclusion, this paper presents a novel recursive updating algorithm for DeePC to efficiently handle computational challenges. The algorithm utilizes SVD for low-dimensional transformations and fast updates. It is flexible, accommodating various data-driven methods that use Pseudoinverse and Hankel matrices, as demonstrated through a comparison to Subspace Predictive Control.

 \appendix

\subsection{
\change{Proofs of Lemmas~\ref{lemma:deepc_spc1} and~\ref{lemma:deepc_spc2}}
} \label{apenendix:lemma_proof}
First, we derive a relationship from the stochastic LTI system~\eqref{eqn:SLTI}.
By propagating the dynamics from time $t$, we can formulate the next $n_{pred}$-step output as:
\begin{align} \label{eqn:ground_truth1}
    y_{real} = \Gamma x_{t} + K_1 u_{pred} + K_2 e_{pred} 
\end{align}
where $e_{pred} := e_{t:t+n_{pred}-1}$ and
\begin{align*}
    \Gamma = \begin{bmatrix}
        C^{\top} & (CA)^{\top} & (CA^2)^{\top} & \cdots & (CA^{n_{pred}-1)^{\top}}
    \end{bmatrix}^{\top}, 
\end{align*}
\begin{align*}
    K_1 = \begin{bmatrix}
        D & 0& 0& \cdots & 0 \\
        CB & D & 0 & \cdots & 0 \\
        CAB & CB & D & \cdots & 0 \\
        \cdots & \cdots & \ddots &\ddots & \vdots \\
        CA^{n_{pred}-2}B & CA^{n_{pred}-3}B &\cdots & CB & D
    \end{bmatrix}, 
\end{align*}
\begin{align*}
    K_2 = \begin{bmatrix}
        I & 0& 0& \cdots & 0 \\
        CK & I & 0 & \cdots & 0 \\
        CAK & CB & I & \cdots & 0 \\
        \cdots & \cdots & \ddots &\ddots & \vdots \\
        CA^{n_{pred}-2}K & CA^{n_{pred}-3}K &\cdots & CK & I
    \end{bmatrix}
\end{align*}
By replacing $e_t = y_t - Cx_t - Du_t$ in the state propagation in~\eqref{eqn:SLTI}, a predictor-form state-space model can be formulated as: $x_{t+1} = \tilde{A}x_t + \tilde{B}u_t + Ky_t, \; y_t = Cx_t + Du_t + e_t$, where $\tilde{A}=A-KC$ and $\tilde{B}=B-KD$. From this model, we can find a relation between $x_t$ and $x_{t-n_{init}}$ by:
\begin{align*}
    x_t = \tilde{A}^P x_{t-n_{init}} + K_3 u_{init} + K_4 y_{init}
\end{align*}
By replacing the above equation in~\eqref{eqn:ground_truth1}, we can find:
\begin{equation} \label{eqn:yreal2}
\begin{aligned}
    & y_{real}  =  \begin{bmatrix}
        \Gamma K_4 & \Gamma K_3 & K_1 
    \end{bmatrix}
    \begin{bmatrix}
        y_{init}\\u_{init}\\u_{pred}
    \end{bmatrix} \\ & \quad 
     + \tilde{A}^P x_{t-n_{init}} + K_2 e_{pred} 
\end{aligned}
\end{equation}
The above linear relation can be extended to the Hankel matrices $H_u, H_y$ by
\begin{equation} \label{eqn:Hankel_linear}
\begin{aligned}
    & H_{y,pred}  =  
    \begin{bmatrix}
        \Gamma K_4 & \Gamma K_3 & K_1  
    \end{bmatrix}
    \begin{bmatrix}
        H_{y,init}\\H_{u}
    \end{bmatrix}
    \\ & \quad + \tilde{A}^P X_{-n_{init}} + K_2 H_{e,pred} 
\end{aligned}
\end{equation}
where $H_{e,pred}$ represents the prediction part in $H_e$, similar to the definition of $H_{y,pred}$. Besides, $X_{-n_{init}}:= \begin{bmatrix}
        x_{1-n_{init}} & x_{2-n_{init}} & \cdots & x_{T-L+1-n_{init}}
    \end{bmatrix}$, where the time corresponds to that of $\{u_i\}_{i=1}^T$ and $\{y_i\}_{i=1}^T$ for constructing $H_u$ and $H_y$.

\begin{proof} [For Lemma~\ref{lemma:deepc_spc1}]
The data-driven prediction~\eqref{eqn:bilevel_prediction} can be written in the explicit form: $  y_{pred} = H_{y,pred} Z^{\dag}\begin{bmatrix}
        y_{init}\\u_{init}\\u_{pred}
    \end{bmatrix}$ by defining $Z:=\begin{bmatrix}
        H_{y,init}\\H_{u}
    \end{bmatrix}$ for simplicity.
It can be rewritten as
\begin{align} \label{eqn:Hankel_pred}
    y_{pred} = \frac{1}{T}H_{y,pred} Z^{\top}( \frac{1}{T} Z Z^{\top})^{-1}
    \begin{bmatrix}
        y_{init}\\u_{init}\\u_{pred}
    \end{bmatrix}
\end{align}
under Assumption~\ref{assmp:H_full_rank},  
which almost surely ensures that the inverse exists for stochastic LTI systems. Besides, Assumption~\ref{assmp:input_infnity} ensures the existence of matrix correlation in~\eqref{eqn:Hankel_pred}.

By replacing $H_{y,pred}$ by~\eqref{eqn:Hankel_linear} to~\eqref{eqn:Hankel_pred}, we can find the following result:
\begin{equation} \label{eqn:Hankel_vanish}
\begin{aligned}
    & \lim_{T \rightarrow \infty} \frac{1}{T}H_{y,pred} Z^{\top}( \frac{1}{T} Z Z^{\top})^{-1} \\
    =& \begin{bmatrix}
        \Gamma K_3 & K_1 &\Gamma K_4 
    \end{bmatrix} + \\
    &\lim_{T \rightarrow \infty} \frac{1}{T}(\tilde{A}^P X_{-n_{init}} + K_2 H_{e,pred})Z^{\top}( \frac{1}{T} Z Z^{\top})^{-1} \\
    =&\begin{bmatrix}
        \Gamma K_3 & K_1 &\Gamma K_4 
    \end{bmatrix}
\end{aligned}
\end{equation}
where the latter term in the second equation vanishes due to Assumption~\ref{assmp:Ak_stable} and the lack of correlation between the $u_t$ and $e_t$ in open-loop data. Referring to~\eqref{eqn:yreal2}, and~\eqref{eqn:Hankel_vanish} and Assumption~\ref{assmp:Ak_stable}, we can demonstrate consistency in the prediction made by~\eqref{eqn:Hankel_pred} as $T \rightarrow \infty$ by:
\begin{align*}
    \mathbb{E}_{e} (y_{pred}-y_{real}) = \mathbb{E}_{e} (\tilde{A}^P x_{t-n_{init}} + K_2 e_{pred}) = \mathbf{0}
\end{align*}

\end{proof}

\begin{proof}
 [For Lemma~\ref{lemma:deepc_spc2}] The proof is very similar to the one for For Lemma~\ref{lemma:deepc_spc1}. The only difference is that $\lim_{T \rightarrow \infty} \frac{1}{T} K_2 H_{e,pred})Z^{\top} \neq 0$ in general
 for closed-loop data. However, under the specific setup, i.e. $D=0$ in the LTI system or the I/O data is collected by feedback control with at least one sample time delay, we again have $\lim_{T \rightarrow \infty} \frac{1}{T} K_2 H_{e,pred})Z^{\top} = 0$.
\end{proof}

\subsection{Data-driven prediction: a specific form} \label{apenendix:B}
This appendix explains how to transform the data-driven prediction in~\eqref{eqn:bilevel_prediction}, \eqref{eqn:spc},  and~\eqref{eqn:bilevel_prediction_cl} into the specific form~\eqref{eqn:ypred_spec}. In addition, the ground truth is derived in the form of~\eqref{eqn:ypred_spec}.

For~\eqref{eqn:bilevel_prediction} and~\eqref{eqn:spc}, the result directly comes from its explicit solution, given in~\eqref{eqn:Hankel_pred}. For~\eqref{eqn:bilevel_prediction_cl}, each output prediction $y_{pred}(i)$ is firstly replaced by the explicit solution of~\eqref{eqn:Hankel_pred} with $n_{pred}=1$. After that, each $y_{pred}(i)$ can be reformulated in the form of~\eqref{eqn:ypred_spec} by dynamic programming from $i=1$.

Next, we explain the derivation of the ground truth. 
The ground truth in the form of~\eqref{eqn:ypred_spec} is designed by choosing: $K_{y,init} = \Gamma K_4, K_{u,init} = \Gamma K_3, K_{u,pred} = K_1$ from~\eqref{eqn:yreal2}.
By~\eqref{eqn:yreal2} and Assumption~\ref{assmp:Ak_stable}, it is trivial to prove that it generates consistent prediction.



\subsection{Integration of forgetting factors} \label{apenendix:C}
This appendix introduces that how to extend the recursive DeePC (Algorithm~\ref{alg:Rdeepc}) to two adaptive DeePC algorithms, and describes their equivalent efficient methods similar to Algorithm~\ref{alg:ERdeepc}.

First, we consider one two-step adaptive updating method for the Hankel matrix by:
    \begin{align*} 
        & H \leftarrow \alpha H \\
        & H \leftarrow 
        \begin{bmatrix}
        H &
        \begin{bmatrix}
        y_{t-L:t-1}\\ u_{t-L:t-1}
        \end{bmatrix}      
        \end{bmatrix}  
    \end{align*}
In the first step, a constant $\alpha < 1$ is chosen as a forgetting factor, allowing outdated data to be gradually phased out. This leads to the first adaptive DeePC, which is achieved by replacing the recursive update in the step 1) of Algorithm~\ref{alg:Rdeepc} with the mentioned update method. 
In addition, an efficient version can be implemented using Algorithm~\ref{alg:ERdeepc} by substituting the SVD update in step 1) with Algorithm~\ref{alg:SVD2}. It leverages the fact that $\alpha H = \alpha U_1\Sigma V_1 =  U_1\alpha\Sigma V_1$. Its computational complexity is $\mathcal{O}(\textit{row}_{H}r_{H}^2)$ and requires $\mathcal{O}(\textit{row}_{H}r_{H})$ space. 

\begin{algorithm} 
  \caption{Fast SVD updating with a forgetting factor} \label{alg:SVD2}
{
\textbf{Given:} Current SVD components: $U_1, \Sigma$
\begin{itemize} [nolistsep,leftmargin=1.5em]
    \item[1)] Update $\Sigma$ by $\Sigma \leftarrow \alpha \Sigma $
    
    \item[2)] Conduct steps 1), 2), 3) in Algorithm~\ref{alg:SVD}. 
\end{itemize}
}
\end{algorithm}

Next, we explore another two-step adaptive updating method for the Hankel matrix:
\begin{equation} \label{eqn:downdate}
    \begin{aligned} 
        & H \leftarrow H(:, 2:end) \\
        & H \leftarrow 
        \begin{bmatrix}
        H &
        \begin{bmatrix}
        y_{t-L:t-1}\\ u_{t-L:t-1}
        \end{bmatrix}      
        \end{bmatrix}  
    \end{aligned}
\end{equation}
The first step  involves a  matrix downdating operation, removing the first column from the stacked Hankel matrix $H$. This allows us to formulate the second adaptive DeePC by updating step 1) of Algorithm~\ref{alg:Rdeepc} with the method described above. Moreover, an efficient variant can be achieved by updating Algorithm~\ref{alg:ERdeepc} at step 1) with Algorithm~\ref{alg:SVD3}. A minor adjustment involves maintaining $U_1$ with $\textit{row}_{H}$ columns, expressed as $ H = \begin{bmatrix}U_1 & U_2\end{bmatrix} \begin{bmatrix}
        \Sigma & \mathbf{0} 
    \end{bmatrix}
    \begin{bmatrix}V_1 & V_2\end{bmatrix}^{\top} 
    = U_1 \Sigma V_1^{\top} $.
This approach has a computational complexity of $\mathcal{O}(\textit{row}_{H}^3)$ and requires $\mathcal{O}(\textit{row}_{H}\textit{col}_{H})$ space as $H$ is recorded. 

\begin{algorithm} 
  \caption{Fast SVD downdating and updating} \label{alg:SVD3}
{
\textbf{Given:} Current SVD components: $U_1, \Sigma$ and current
\begin{itemize} [nolistsep,leftmargin=1.5em]
    \item[1)] Retrieve the first column of $H$ as $b$. Retrieve \change{the column $a$ to be added to} $H$, i.e. $\begin{bmatrix}
        y_{t-L:t-1}\\ u_{t-L:t-1}
        \end{bmatrix}$ at time $t$. Update $H$ by~\eqref{eqn:downdate}.
    \item[2)]
    Update $U_1, \; \Sigma$ by~\cite{gu1995downdating}:
    \begin{itemize} [nolistsep,leftmargin=1.0em]
        \item[] Compute $z_1 = U_1^{\top}b$
        \item[] Compute eigendecomposition of $\Sigma^{2} - z_1z_1^{\top}$: $C_1\bar{\Sigma}_1C_1^{\top}$
        \item[] Update $U_1 \leftarrow U_1C_1$, $\Sigma \leftarrow \sqrt{\bar{\Sigma}_1}$
    \end{itemize}     
    \item[3)] Update $U_1, \; \Sigma$ by~\cite{bunch1978updating}:
    \begin{itemize} [nolistsep,leftmargin=1.0em]
        \item[] Compute $z_2 = U_1^{\top}a$
        \item[] Compute eigendecomposition of $\Sigma^{2} + z_2z_2^{\top}$: $C_2\bar{\Sigma}_2C_2^{\top}$
        \item[] Update $U_1 \leftarrow U_1C_2$, $\Sigma \leftarrow \sqrt{\bar{\Sigma}_2}$
    \end{itemize} 
\end{itemize}
}
\end{algorithm}

If the DeePC in the form of Problem 1 is used in each adaptive DeePC algorithm,
this algorithm and its aforementioned efficient version are equivalent. The proof of this equivalence closely mirrors that of Lemma~\ref{lemma:alg_equal}.

\changeE{
 \subsection{Equivalent formulations for Pseudoinverse-based output prediction} \label{apenendix:D}
 }
 

This appendix initially outlines two optimization problems that yield the same solution as a Pseudoinverse-based least squares solution described in Lemma~\ref{lemma:pseu1}. Following this, Corollary~\ref{lemma:pseu2} helps to derive two equivalent formulations that achieve the same outcome as the Pseudoinverse-based output prediction in~\eqref{eqn:pse_deepc}. Lastly, Lemma~\ref{lemma:pseu3} demonstrates that these formulations can be translated into an optimization problem in the form of Problem 1, thus confirming the existence of their equivalent low-dimensional versions.

\begin{lemma} \label{lemma:pseu1}
For any matrix $M$ and any vector $b$, the following two optimization problems,
\begin{equation} \label{eqn:app_pseu1}
\begin{aligned}
    x^{\ast} & = \argmin_{x} \lVert x \rVert_2^2 \\
    & \text{s.t.}\; M^TMx = M^Tb
\end{aligned}
\end{equation}
\begin{equation} \label{eqn:app_pseu2}
\begin{aligned}
    & x^{\ast} = \argmin_{x} \lim_{\lambda \rightarrow 0_+} \lambda\lVert x \rVert_2^2 + \lVert Mx-b \rVert_2^2 
\end{aligned}
\end{equation}
are equivalent with the same unique optimal solution, 
    \begin{align*}
    x^{\ast}=M^{\dag}b.
    \end{align*}
\end{lemma}
\begin{proof}
    The Pseudoinverse provides a unique solution with the minimal L2 norm to least squares problems~\cite{penrose1956best}, formulated as:
    \begin{align*}
    x^{\ast}=M^{\dag}b &  = \argmin_{x} \lVert x \rVert_2^2 \\
    & \text{s.t.}\; x = \argmin_{x_l} \lVert Mx_l - b \rVert_2^2 
    \end{align*}
    Transform the convex lower-level least squares by KKT conditions~\cite{boyd2004convex}, it is equivalent to the problem~\eqref{eqn:app_pseu1}.
    
    Regarding the problem~\eqref{eqn:app_pseu2}, its optimal solution is $x^{\ast} = \lim_{\lambda \rightarrow 0_+} (A^{T}A+\lambda I)^{-1}A^{T}b$ 
    by KKT conditions.
    Then the truth that $M^{\dag} = \lim_{\lambda \rightarrow 0_+} (A^{T}A+\lambda I)^{-1}A^{T}$ \cite[Theorem 4.3]{barata2012moore} completes the proof.
\end{proof}

\begin{corollary} \label{lemma:pseu2}
For any given vector $u_{init},  y_{init}, u_{pred}$ and matrix $M_1, M_2$ with proper size, the following two optimization problems,
\begin{equation} \label{eqn:app_pseu_y1}
\begin{aligned}
    y_{pred} & = M_2 g \\
    g & =  \argmin_{ g_l} \lVert g_l \rVert_2^2 \\
    & \quad \text{s.t.}\; M_1^T M_1 g_l = M_1^T \begin{bmatrix}
        y_{init}\\u_{init}\\u_{pred}
    \end{bmatrix} 
\end{aligned}
\end{equation}
\begin{equation} \label{eqn:app_pseu_y2}
\begin{aligned}
    y_{pred} & = M_2 g \\
    g & =  \argmin_{ g_l} \lim_{\lambda \rightarrow 0_+} \lambda\lVert g_l \rVert_2^2 + \lVert \sigma  \rVert_2^2 \\
    & \quad \text{s.t.} \; M_1 g_l = \begin{bmatrix}
        y_{init}\\u_{init}\\u_{pred}
    \end{bmatrix} + \sigma
\end{aligned}    
\end{equation}
are equivalent and compute
the same unique optimal solution, formulated as:
\begin{equation} \label{eqn:app_pseu_y3}
\begin{aligned}
   y_{pred} &= M_2 g\\
    g &= M_1^{\dag}\begin{bmatrix}
        y_{init}\\u_{init}\\u_{pred}
    \end{bmatrix} 
\end{aligned}    
\end{equation}
\end{corollary}
\begin{proof}
    Based on Lemma~\ref{lemma:pseu1}, both the optimization problems in~\eqref{eqn:app_pseu_y1} and~\eqref{eqn:app_pseu_y2} have the same optimal solution $g$ and therefore the same prediction $y_{pred}$ formulated in~\eqref{eqn:app_pseu_y3}.
\end{proof}

Based on Corollary~\ref{lemma:pseu2}, by choosing $M_1 = H_{y,pred}, M_2 = \begin{bmatrix}H_{y,init}^T & H_{u}^T\end{bmatrix}^T$, the following two output prediction methods: 
\begin{equation} \label{eqn:pse_deepc_1}
\begin{aligned}
y_{pred} & = H_{y,pred} g \\
    g & =  \argmin_{ g_l} \lVert g_l \rVert_2^2 \\
     \text{s.t.} &\; \begin{bmatrix}
        H_{y,init}\\H_{u}
    \end{bmatrix}^T \begin{bmatrix}
        H_{y,init}\\H_{u}
    \end{bmatrix} g_l = \begin{bmatrix}
        H_{y,init}\\H_{u}
    \end{bmatrix}^T \begin{bmatrix}
        y_{init}\\u_{init}\\u_{pred}
    \end{bmatrix} 
\end{aligned}    
\end{equation}
\begin{equation} \label{eqn:pse_deepc_2}
\begin{aligned}
    y_{pred} & = H_{y,pred} g \\
    g &=  \argmin_{ g_l} \lim_{\lambda \rightarrow 0_+} \lambda\lVert g_l \rVert_2^2 + \lVert \sigma  \rVert_2^2 \\
    & \text{s.t.} \; \begin{bmatrix}
        H_{y,init}\\H_{u}
    \end{bmatrix} g_l = \begin{bmatrix}
        y_{init}\\u_{init}\\u_{pred}
    \end{bmatrix} + \sigma
\end{aligned}    
\end{equation}
leads to the same results as the Pseudoinverse-based output
prediction in~\eqref{eqn:pse_deepc}

Lastly, we will illustrate some low-dimensional versions for~\eqref{eqn:pse_deepc_1}, \eqref{eqn:pse_deepc_2}, and will prove the equivalence in Lemma~\ref{lemma:pseu3}. Separate the transformed version of the Hankel matrix by $\bar{H}:=\begin{bmatrix}
        \bar{H}_y \\ \bar{H}_u
\end{bmatrix}, \bar{H}_y = \begin{bmatrix} \bar{H}_{y,init}\\\bar{H}_{y,pred} \end{bmatrix}$, similar to the separation operation for the original Hankel matrix.
Based on Lemma~\ref{lemma:pseu2}, by choosing $M_1 = \bar{H}_{y,pred}, M_2 = \begin{bmatrix}\bar{H}_{y,init}^T & \bar{H}_{u}^T\end{bmatrix}^T$, the following two output prediction methods: 
\begin{equation} \label{eqn:pse_deepc_3}
\begin{aligned}
y_{pred} & = \bar{H}_{y,pred} \bar{g} \\
    \bar{g} & =  \argmin_{ \bar{g}_l} \lVert \bar{g}_l \rVert_2^2 \\
     \text{s.t.} &\; \begin{bmatrix}
        \bar{H}_{y,init}\\\bar{H}_{u}
    \end{bmatrix}^T \begin{bmatrix}
        \bar{H}_{y,init}\\ \bar{H}_{u}
    \end{bmatrix} \bar{g}_l = \begin{bmatrix}
        \bar{H}_{y,init}\\\bar{H}_{u}
    \end{bmatrix}^T \begin{bmatrix}
        y_{init}\\u_{init}\\u_{pred}
    \end{bmatrix} 
\end{aligned}    
\end{equation}

\begin{equation} \label{eqn:pse_deepc_4}
\begin{aligned}
    y_{pred} & = \bar{H}_{y,pred} \bar{g} \\
    \bar{g} & =  \argmin_{ \bar{g}_l} \lim_{\lambda \rightarrow 0_+} \lambda\lVert \bar{g}_l \rVert_2^2 + \lVert \sigma  \rVert_2^2 \\
    & \text{s.t.} \; \begin{bmatrix}
        \bar{H}_{y,init}\\\bar{H}_{u}
    \end{bmatrix} \bar{g}_l = \begin{bmatrix}
        y_{init}\\u_{init}\\u_{pred}
    \end{bmatrix} + \sigma
\end{aligned}    
\end{equation}
leads to the same results:
\begin{equation} \label{eqn:pse_deepc_5}
\begin{aligned}
   y_{pred} &= \bar{H}_{y,pred}\bar{g}\\
    \bar{g} &= \begin{bmatrix}
        \bar{H}_{y,init}\\\bar{H}_{u}
    \end{bmatrix}^{\dag}\begin{bmatrix}
        y_{init}\\u_{init}\\u_{pred}
    \end{bmatrix} 
\end{aligned}    
\end{equation}

\begin{lemma} \label{lemma:pseu3}
For any given vector $u_{init},  y_{init}, u_{pred}$ and any Hankel matrix $H$ with its transformed version $\bar{H}$, \eqref{eqn:pse_deepc_1}, \eqref{eqn:pse_deepc_2}, \eqref{eqn:pse_deepc_3}, \eqref{eqn:pse_deepc_4} and \eqref{eqn:pse_deepc_5} have the same Pseudoinverse-based output
prediction $y_{pred}$ given in~\eqref{eqn:pse_deepc}.
\end{lemma}
\begin{proof}
Consider the optimization problem:
\begin{align*}
    (y_{pred}, g) &=  \argmin_{y_{pred,l}, g_l} \lim_{\lambda \rightarrow 0_+} \lambda\lVert g_l \rVert_2^2 + \lVert \sigma  \rVert_2^2 \\
    & \text{s.t.} \; \begin{bmatrix}
        H_{y,init}\\H_{u}
    \end{bmatrix} g_l = \begin{bmatrix}
        y_{init}\\u_{init}\\u_{pred}
    \end{bmatrix} + \sigma \\
    & \quad \;\; y_{pred,l} = H_{y,pred} g
\end{align*}    
The above problem also has an unique optimal solution~\eqref{eqn:pse_deepc},  because this problem and the strictly convex problem~\eqref{eqn:pse_deepc_2} can be transformed into the same linear equations by KKT constraints. By the fact that $\lVert g\rVert_2^2 = g^{\top}\begin{bmatrix}V_1 & V_2\end{bmatrix} \begin{bmatrix}V_1 & V_2\end{bmatrix}^{\top}g = \lVert V_1 g\rVert_2^2 + \lVert V_2 g\rVert_2^2$, this problem is in the form of Problem 1. Its low-dimensional version in the form of Problem 2 is formulated as:
\begin{align*}
    (y_{pred}, \bar{g}) &=  \argmin_{ y_{pred,l}, \bar{g}_l} \lim_{\lambda \rightarrow 0_+} \lambda\lVert \bar{g}_l \rVert_2^2 + \lVert \sigma  \rVert_2^2 \\
    & \text{s.t.} \; \begin{bmatrix}
        \bar{H}_{y,init}\\\bar{H}_{u}
    \end{bmatrix} \bar{g}_l = \begin{bmatrix}
        y_{init}\\u_{init}\\u_{pred}
    \end{bmatrix} + \sigma \\
    & \quad \;\; y_{pred,l} = \bar{H}_{y,pred} g
\end{align*} 
Similarly, the above problem has the same unique optimal solution as~\eqref{eqn:pse_deepc_4}. Therefore, we can conclude that all the methods, \eqref{eqn:pse_deepc_1}, \eqref{eqn:pse_deepc_2}, \eqref{eqn:pse_deepc_3}, \eqref{eqn:pse_deepc_4}, \eqref{eqn:pse_deepc_5}, have the same output
prediction $y_{pred}$ as the one in~\eqref{eqn:pse_deepc} by Lemma~\ref{lemma:deepc_equal} and Corollary~\ref{lemma:pseu2}.
\end{proof}

\changeE{
 \subsection{SPC equivalents and an effcient recursive SPC algorithm} \label{apenendix:E}
 }

Consider a bi-level DeePC
\begin{equation} \label{eqn:pse_bideepc_1}
\begin{aligned}
\min_{\substack{g,u_{pred}\\y_{pred}}} \; & J(u_{pred},y_{pred}) \\
\text{s.t.}\; & y_{pred} \in \mathbb{Y}, u_{pred} \in \mathbb{U}  \\
    & \eqref{eqn:pse_deepc_1} 
\end{aligned}
\end{equation}
, its low-dimensional version,
\begin{equation} \label{eqn:pse_bideepc_3}
\begin{aligned}
\min_{\substack{\bar{g},u_{pred}\\y_{pred}}} \; & J(u_{pred},y_{pred}) \\
\text{s.t.}\; & y_{pred} \in \mathbb{Y}, u_{pred} \in \mathbb{U}  \\
    & \eqref{eqn:pse_deepc_3}
\end{aligned}
\end{equation}
and a low-dimensional SPC:
\begin{equation} \label{eqn:pse_bideepc_2}
\begin{aligned}
\min_{g,\sigma} \; & J(u_{pred},y_{pred}) \\
\text{s.t.}\; & y_{pred} \in \mathbb{Y}, u_{pred} \in \mathbb{U}  \\
    & y_{pred} = K\begin{bmatrix}
            y_{init}\\u_{init}\\u_{pred}
        \end{bmatrix}   \\
    & K = \bar{H}_{y,pred}\begin{bmatrix}
        \bar{H}_{y,init}\\\bar{H}_{u}
    \end{bmatrix}^{\dag}
\end{aligned}
\end{equation}
They are equivalent to the SPC~\eqref{eqn:spc} without any assumption as stated in Lemma~\ref{lemma:appen_spc}.

\begin{lemma} \label{lemma:appen_spc}
SPC~\eqref{eqn:spc}, its low-dimensional version~\eqref{eqn:pse_bideepc_2} and the two bi-level DeePCs~\eqref{eqn:pse_bideepc_1}, \eqref{eqn:pse_bideepc_3} are equivalent.
\end{lemma}
\begin{proof}
    The only difference between the two controllers is their prediction parts. By Lemma~\ref{lemma:pseu3}, all the predictions can be written in the same explicit form finishes the proof: $  y_{pred} = H_{y,pred} \begin{bmatrix}
        H_{y,init}\\H_{u}
    \end{bmatrix}^{\dag}\begin{bmatrix}
        y_{init}\\u_{init}\\u_{pred}
    \end{bmatrix}$
\end{proof}

In addtion, an efficient recursive SPC method is presented in Algorithm~\ref{alg:ERSPC} as an variant of Algorithm~\ref{alg:ERdeepc}.

 \begin{algorithm} 
  \caption{Efficient Recursive SPC}
  \label{alg:ERSPC} 
{
\begin{itemize} [nolistsep,leftmargin=1.5em]
    \item[0)] Retrieve some persistently excited past I/O data. Construct $H$ and compute its SVD. Build the initial low-dimensional SPC controller~\eqref{eqn:pse_bideepc_2} or its equivalent low-dimensional bi-level DeePC~\eqref{eqn:pse_bideepc_3}.
    \item[1)] Retrieve the recent $L$-step measurements. Update the SVD components based on Algorithm~\ref{alg:SVD} and update $\bar{H}$.
    \item[2)] Retrieve the recent $t_{init}$-step measurements. Solve the low-dimensional SPC or its equivalence, and apply the optimal input as $u_t=u_{pred}^{\ast}(1)$.
    \item[3)] Pause until the subsequent sampling time, update $t\leftarrow t+1$ and revert to step 1.
\end{itemize}
}
\end{algorithm}







\bibliographystyle{ieeetr}
\bibliography{ref.bib}

\end{document}